\documentclass[useamsfonts]{pasj00}
\usepackage{graphicx,epsfig} 
\usepackage{amsthm}
\usepackage{amssymb}
\usepackage{enumerate}
\usepackage{xspace}
\usepackage{euscript}
\usepackage{graphicx}
\usepackage{amscd}
\usepackage{epsfig}
\usepackage{tabularx}
\usepackage{mathrsfs}

\newcommand{\drv}[2]{\frac{\partial #1}{\partial #2}   }

\newcommand{\ddrv}[2]{\frac{\partial^{2}  #1}{{\partial #2}^2}   }

\begin{document}
%\hyphenation {Schwarz-schild}
%\hyphenation {Abra-mo-wicz}

\SetRunningHead{Rebusco P.}{Non-linear resonance}
\Received{2003/12/25}%{yyyy/mm/dd}
\Accepted{2004/03/15}

\title{Twin peaks kHz QPOs: mathematics of the $3:2$ orbital resonance}

\author{P.~\textsc{Rebusco}\altaffilmark{1,2}}

\affil{$^1$~Max-Planck-Instutut f{\"u}r Astrophysik, $D-85741$ Garching, Germany}
\affil{$^2$~Department of Physics, Trieste University, I-34\,127 Trieste, Italy}

\KeyWords{ General relativity --
 Accretion  -- X-rays: individual (Sco X-1) -- QPOs}

\maketitle
\begin{abstract}
Using the method of multiple scales, one can derive an analytic
solution that describes the behaviour of weakly coupled, non-linear
oscillations in  nearly Keplerian discs around neutron stars or  black
holes close to the 3:2 orbital epicyclic resonance.  The solution
obtained agrees with the previous numerical simulation. Such result
may be relevant to the kilohertz quasi-periodic variability in X-ray
fluxes observed in many Galactic black hole and neutron star sources. With a particular choice of  tunable parameters, the solution fits accurately the observational data for Sco X-1. 
\end{abstract}

\section{Introduction}

Many Galactic black hole and neutron star sources in low  mass X-ray
binaries show both chaotic and quasi-periodic variability in their
observed X-ray fluxes. Some of the quasi-periodic oscillations (QPOs)
are in the kHz range and often come in pairs $(\nu_{\rm upp}, \nu_{\rm
down})$ of {\it twin peaks} in the Fourier power spectra (e.g., van der
Klis, 2000). In all four black hole sources with twin peak kHz QPOs
pairs, $\nu_{\rm upp}/\nu_{\rm down} = 3/2\,$ (McClintock \&
Remillard, 2003). For  neutron stars sources the ratio of twin-peak
frequencies is mostly  close to $3/2$ too
 (Abramowicz, Bulik, Bursa, Klu\'zniak, 2003). Based
on these and other properties of QPOs Klu\'zniak \& Abramowicz (2000)
concluded that twin peak kHz QPOs are due to a resonance in the
accretion disc oscillation modes and they noticed that the specific $3:2$ ratio could be a consequence of strong gravity.

According to the standard Shakura-Sunyaev accretion disc model, matter
spirals down into the central black hole along stream lines that are located almost on the equatorial plane $\theta = \theta_0 = \pi/2$, and that locally differ only slightly from a family of concentric circles $r = r_0 = const$. The small deviations, $\delta r = r - r_0$, $\delta \theta = \theta - \theta_0$ are governed, with accuracy to linear terms, by

%======================================================================
%==
%==        Equation 1
%==
%======================================================================

\begin{equation}
\label{Equation1}
\delta \ddot r + \omega_r^2 \,\delta r = \delta a_r ,
~~~~
\delta \ddot \theta + \omega_{\theta}^2\,\delta \theta = \delta a_{\theta} .
\end{equation}

\noindent Here the dot denotes the time derivative. For purely Keplerian (free) motion $\delta a_r = 0$, $\delta a_{\theta} = 0$ and the above equations describe two uncoupled harmonic oscillators with the eigenfrequencies $\omega_{\theta}$, $\omega_r$.

\begin{figure}[!tb]
\begin{center}
\FigureFile(0.48\textwidth,0.48\textwidth){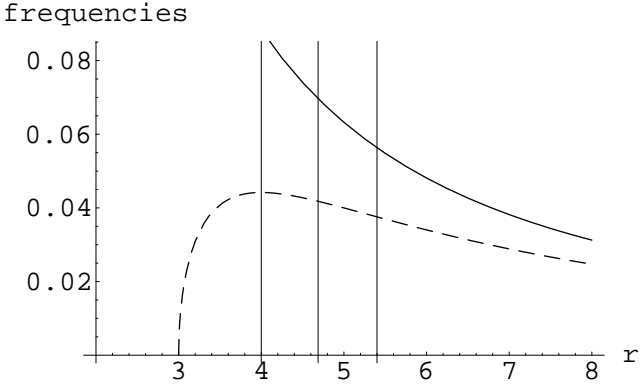}
\end{center}
\caption{Epicyclic  frequencies in Schwarzschild's metric:
meridional (solid line) and radial one (dashed line).
The radius is in units of $r_G$. The vertical lines indicates
the radii ($r_{3:2}$,$r_{5:3}$,$r_{4:2}$) where the ratio 
between the two frequencies is 
$3:2$ , $5:3$ and $4:2$ (from right to left).}
\label{fig1}
\end{figure}

\noindent We shall start with an argument which appeals to physical
intuition and which shows that the resonance now discussed is a very
natural, indeed necessary, consequence of strong gravity. Assume that
in thin discs, random fluctuations have $\delta r \gg \delta
\theta.$\footnote{In actual calculations this additional assumption is
not made.}  Thus, $\delta r \delta \theta$ is a first order term in $\delta \theta$ and should be included in the first order equation for vertical oscillations (\ref{Equation1}). The equation now takes the form,

%======================================================================
%==
%==        Equation 2
%==
%======================================================================

\begin{equation}
\label{Equation5}
\delta \ddot \theta + \omega_{\theta}^2\left [ 1 + h\,\delta r \right ] \delta \theta = \delta a_{\theta},
\end{equation}

\noindent where $h$ is a known constant. The first order equation for $\delta r$ has the solution $\delta r = A_0 \cos (\omega_r \,t)$. Inserting this in (\ref{Equation5}) together with $\delta a_{\theta} = 0$, one arrives to the Mathieu equation ($A_0$ is absorbed in $h$), 

%======================================================================
%==
%==        Equation 3
%==
%======================================================================

\begin{equation}
\delta \ddot \theta + \omega_{\theta}^2\left [ 1 + h \,\cos (\omega_r \,t)\right ] \delta \theta = 0, 
\end{equation}

\noindent that describes the {\it parametric resonance}. From the theory of the Mathieu equation one knows that when

%======================================================================
%==
%==        Equation 4
%==
%======================================================================

\begin{equation}
{\omega_r \over \omega_{\theta}} = {\nu_r \over \nu_{\theta}} = {2 \over n}, ~~~~n =1,\,2, \,3 ..., 
\end{equation}

\noindent the parametric  resonance is excited (Landau \& Lifshitz,
1976). The resonance is strongests for the smallest possible value of
$n$. Because near black holes and neutron stars $\nu_r < \nu_{\theta}$
(see Figure \ref{fig1}), the smallest possible value for resonance is
$n = 3$, which means that $2\,\nu_{\theta} = 3\,\nu_r$. This is an
example of a situation in which parametric resonance works in a thin
accretion disc and it intuitively explains the observed 3:2 ratio
(Klu\'zniak and Abramowicz, 2002), because, obviously,

%======================================================================
%==
%==        Equation 5
%==
%======================================================================
\begin{equation}
\nu_{\rm upp} = \nu_{\theta},~~~~\nu_{\rm down} = \nu_r.
\end{equation}

\noindent Of course, in real discs neither $\delta r = A_0 \cos
(\omega_r \,t)$, nor $\delta a_{\theta} = 0$ exactly; moreover in
realistic
situations, for purely geodesic motion ($\delta a_{\theta}=\delta
a_r=0$)
 the system does not show increasing amplitudes (the higher terms
prevent this from occurring).\\
 However, one may expect that because these equations are approximately
obeyed in thin discs, the parametric resonance will indeed be excited.
Such a resonance  was found in numerical simulations of oscillations in a nearly Keplerian accretion disc by Abramowicz et al. (2003).
%^^^^^^^^^^^^^^^^^^^^^^^^^^^^^^^^^^^^^^^^^^^^^^^^^^^^^^^^^^^^^^^^^^^^^^^
%^^^^^^^^^^^^^^^^^^^^^^^^^^^^^^^^^^^^^^^^^^^^^^^^^^^^^^^^^^^^^^^^^^^^^^

\section{Nearly-geodesic motion}

The equations roughly outlined in the previous Section will now be derived. 

The equations of the geodesic motion of a unit mass test-particle in
spherical coordinates can be written  in the form ($\tilde{\tau}= \tau/r_G$):

\begin{eqnarray}
\ddot\theta(\tau) & + & \frac{E^2}{2 g_{\theta\theta}}
 \drv{U_{eff}(r,\theta)}{\theta}+2 \Gamma_{r \theta}^{\theta}\dot r(\tau) \dot\theta(\tau)=0 \\
\ddot r(\tau)& + &\frac{E^2}{2 g_{rr}}
 \drv{U_{eff}(r,\theta)}{r}+ \Gamma_{\theta \theta}^{r} {\dot\theta(\tau)}^2+\Gamma_{rr}^{r} {\dot r(\tau)}^2=0 \nonumber \\
\ddot\phi(\tau) &= &0\:\:\:\:, \nonumber 
\label{eqGR}
\end{eqnarray}

\noindent where $U_{eff} = g^{tt} +l g^{t \phi}+l^2 g^{\phi \phi}$ is the effective potential  and $E=-u_t$ is the energy.

Consider a  geodesic on the equatorial plane 
$(r_0,\theta_0=\pi/2,\phi=\Omega \tilde{\tau})$ and perturb it slightly, keeping
 constant  the angular momentum.

\begin{eqnarray}
r(\tilde{\tau}) &=& r_0 + \rho(\tilde{\tau})\:\:\:\:\: \rho<<r_0 \\
\theta(\tilde{\tau}) &=& \frac{\pi}{2}+z(\tilde{\tau})\:\:\:\:\: z<<\pi/2\:\:\:\:\:.\nonumber\\
\end{eqnarray}
Here $\rho(\tilde{\tau})$ and $z(\tilde{\tau})$ denote small devations from
the circular orbit, and  to first order they describe  two uncoupled 
harmonic oscillators with epicyclic eigenfrequencies:

\begin{equation}
\omega_{\theta}^2 = 
\left (\frac{1}{g_{\theta\theta}} \ddrv{U_{eff}}{\theta}\right )_{\!\ell,r_0,\pi/2},
\quad
\omega_r^2 =\left (\frac{1}{g_{rr}} \ddrv{U_{eff}}{r}\right )_{\!\ell,r_0,\pi/2}\:\:\:.
\end{equation}

The ratio of the two frequencies is

\begin{equation}
\frac{\omega_{\theta}^2}{\omega_r^2}=\frac{r_0}{r_0-3}>1\:\:\:\:\:.
\label{rapportoGR}
\end{equation}

This does not happen in the Newtonian case where the radial and vertical
frequencies are always equal to the orbital frequency ($\Omega=GM/r_0^3$).
When continuing to higher orders in the Taylor expansion, the two motions become coupled, and this is why  the 
so-called parametric resonance can arise.
The perturbed equations are nonlinear differential equations with constant coefficients
which are the values of 
the different derivatives of the effective potential $U_{eff}$ and the
metric at equilibrium (see Appendix \ref{subsec:notation}):

\begin{eqnarray}
\label{eqperturbateGR}
\delta a_{\theta} &=& \ddot z(\tilde{\tau})+\omega^2_{\theta}
z(\tilde{\tau}) - f(\rho(\tilde{\tau}),z(\tilde{\tau}),r_0,\theta_0)\\
\delta a_r &=& \ddot\rho(\tilde{\tau})+\omega^2_{r} \rho(\tilde{\tau}) - g(\rho(\tilde{\tau}),z(\tilde{\tau}),r_0,\theta_0)\:\:\:\:.\nonumber
\end{eqnarray}

In the case of  Schwarzschild's metric,

\begin{eqnarray}
f(\rho,z,r_0,\theta_0) &=& c_{11} z \rho + c_b \dot{z} \dot{\rho} +\\
         & &+ c_{21} \rho^2 z + c_{1b} \rho \dot{z} \dot{\rho} +
         c_{03} z^3 \nonumber \\
g(\rho,z,r_0,\theta_0)&=& e_{02} z^2 + e_{20} \rho^2 + e_{z2}
         \dot{z}^2 +\nonumber \\
& &+ e_{30} \rho^3 + e_{1ze2} \rho \dot{z}^2 + e_{12} \rho z^2 +
         \nonumber \\
& & e_{r2} \dot{\rho} ^2  + e_{1re2} \dot{\rho}^2 \rho \:\:\:\:.\nonumber
\end{eqnarray}
 The third order terms in the Taylor expansion provide the damping which saturates the increasing
amplitudes: this motivates stopping the expansion at this order of the perturbation.\\
In first approximation we used Paczy{\'n}ski \& Wiita (1980) pseudo-Newtonian
model.
The perturbed  equations in the model and in general relativity look the same, 
apart from those terms in $g(\rho,z,r_0,\theta_0) $ which involve $\dot{\rho}^2$
 that here are absent.\\
On the geodesics $\delta a_{\theta} = \delta a_r=0$; however,
in  order to explore the features of the  resonance, a  small
additional 
isotropic force,
parametrized by $|\alpha| \in [ 0,1 ]$, can be added (\cite{nostro}); this
force at the moment has no physical meaning (it may be a coupling due for example
to pressure or viscosity), it is just a mathematical tool. Then

\begin{eqnarray}
\delta a_{\theta} &=&- \alpha (c_{21}\rho^2 + c_{03}z^2) z \\
\delta a_r &=&- \alpha (e_{z2} + e_{1ze2}\rho)\dot z^2\:\:\:\:.\nonumber
\end{eqnarray}

When $|\alpha|=0$ the previous equations are just the approximation
of the geodesics, while when  $|\alpha|$ increases, they describe slight
deviations from the free motion.

%^^^^^^^^^^^^^^^^^^^^^^^^^^^^^^^^^^^^^^^^^^^^^^^^^^^^^^^^^^^^^^^^^^^^^^^
\begin{figure}[!tb]
\begin{center}
\FigureFile(100mm,50mm){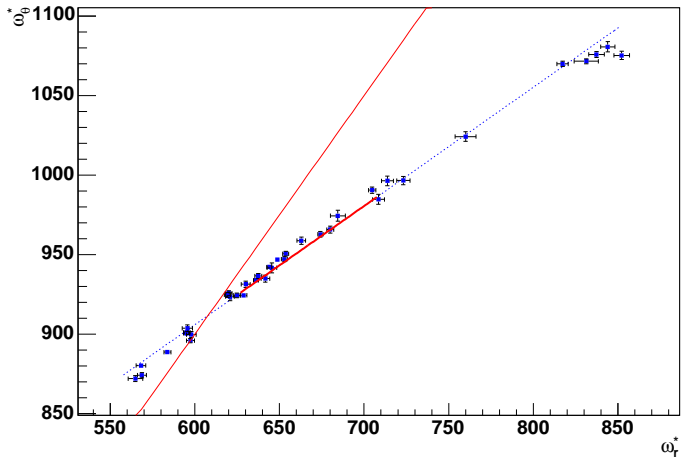}
\end{center}
\caption{The dashed line is the least-squares best-fit to the data
points (the  observed kHz QPOs  frequencies in Sco $X-1$); the thin solid line corresponds to a slope
of $3:2$ (for reference) . The thick segment is the analytic approximation, in which
 the frequencies are scaled for comparison with observations. } 
\label{fig3}
\end{figure}
%^^^^^^^^^^^^^^^^^^^^^^^^^^^^^^^^^^^^^^^^^^^^^^^^^^^^^^^^^^^^^^^^^^^^^^^

\section{The method of multiple scales}

This method is a variation of the straightforward expansion which
leads to a unformly valid approximate solution of systems of weakly
nonlinear differential equations;
the underlying idea is to consider the expansion which
represents the approximate solution to be a function of multiple
independent variables, or scales, instead of a single variable.
The new ``time-like'' independent variables are defined:
$T_k = \epsilon^k t\:\: \:\mbox{for}\:\:\: k=0,1,2...$
Expressing the solution as a function of more variables, treated as
 independent, is an artifice to remove the secular terms, which would
make the solution to explode unphysically.\\
 By writing $T_k$ one makes a formal assumption  of physical slow variation explicit. Indeed $T_k$
define
progressively longer time scales, which are not negligible when $t$ is of the
order of $1/\epsilon^k$ or longer. The characteristic time scale of the
orbital motion near a neutron star or a stellar mass black hole is short 
($T_0\sim 1$ms): hence for $\epsilon$ of the order of $10^{-1}$ the dynamical
effects on the scale $T_3$ become important in the interval of $1$s.\\
The conditions to eliminate the secular terms  give a simpler system of
differential equations: for a set of initial conditions the solution to
such a system is a sum of trigonometric functions. 
The same solution can be found 
by assuming a priori that the asymptotic
 solution is the sum of trigonometric functions in which the
dominant terms have frequencies not far from the original
eigenfrequencies of the uncoupled harmonic oscillators (method of Lindstedt-Poincar\'e). The
 corrections
to the eigenfrequencies depend on the amplitudes and on the
constant coefficients Hence in a Fourier transform
of the geodesics perturbed in the vicinity of $r_{3:2}=27/5$, 
two peaks centered near $3:2$ would be found,
the position of which
 and the ratio depend on the perturbation ($\alpha$) and
on the initial amplitudes.

This method was used to find the approximate solution to equations
(\ref{eqperturbateGR}) (both when $\alpha=0$, hence $\delta
\alpha_{\theta}=\delta \alpha_r=0$ and not):
\begin{eqnarray}
r(t) &=& 27/5 + \epsilon a \cos{(\omega_r^* )t} +\\
     &+&  (\epsilon^2 v+\epsilon^4 v_1) \cos{( 2 \omega_r^*
     )t}+ \nonumber\\ 
&+&(\epsilon^2 p+\epsilon^4 p_1) \cos{(2 \omega_{\theta}^*)t}+
(\epsilon^2 c+ \epsilon^4 c_1)+ \nonumber \\
     &+&  \epsilon^3 d \cos{(\omega_r^* -2
     \omega_{\theta}^*)t}+ \nonumber \\ 
&+& \epsilon^3 e \cos{(\omega_r^* +2 \omega_{\theta}^*)t}+ \epsilon^3 f \cos{(3 \omega_r^*)t}+ \nonumber \\
 &+& \epsilon^4 r \cos{(2 \omega_r^* +2
     \omega_{\theta}^*)t}+ \epsilon^4 s \cos{(4
     \omega_r^*)t}+ \nonumber\\ 
&+&  \epsilon^4 u \cos{(4 \omega_{\theta}^*)t} \nonumber \\
\theta(t) &=& \pi/2 + \epsilon g \cos{(\omega_{\theta}^* )t} + \nonumber\\      &+& (\epsilon^2 h+\epsilon^4 h_1) \cos{( \omega_r^*
     -\omega_{\theta}^*)t}+ \nonumber \\
&+& (\epsilon^2 l +\epsilon^4 l_1)\cos{( \omega_r^*
     +\omega_{\theta}^*)t}+ \epsilon^3 m \cos{(2 \omega_r^*
     - \omega_{\theta}^*)t}+ \nonumber \\ 
&+& \epsilon^3 n \cos{(2 \omega_r^* + \omega_{\theta}^*)t}+ \epsilon^3 o \cos{(3 \omega_{\theta}^*)t}+ \nonumber \\
&+& \epsilon^4 j \cos{(3 \omega_r^* + \omega_{\theta}^*)t}+ 
\epsilon^4 w \cos{(\omega_r^* +
     3 \omega_{\theta}^*)t}+\nonumber \\
&+& \epsilon^4 q \cos{ (\omega_r^* - 3 \omega_{\theta}^*)t}\nonumber
\end{eqnarray}            

The coefficients of the trigonometric functions ($a$,$g$,$c$..) are constants which depend on $r_0$ and the
initial conditions. For simplicity of notation here the  initial
phases are equal to zero. 

\section{Symmetry and regions of resonance}
The Taylor expansion at constant angular momentum to third order near the stationary 
point of the effective potential leads to two nonlinear differential
 equations with constant coefficients:

\begin{eqnarray}
\ddot z(t)+\omega_{\theta}^2 z(t) &=& \sum_{i,j,k,l}a_{i j k l}z(t)^i \rho(t)^j
\dot{z(t)^k} \dot{\rho(t)^l}\\
\ddot r(t)+\omega_{r}^2 r(t) &=& \sum_{q,r,o,p}b_{q r o p}z(t)^q \rho(t)^r
\dot{z(t)^o} \dot{\rho(t)^p}\:\:\:\:\:.\nonumber
\end{eqnarray}
The polynomials of third degree ($i+j+k+l=3\:\: \forall i,j,k,l$) which 
constitute the nonlinear part of
the equation depend on the symmetry of the effective potential and of the metric.\\

 The first part of the method of multiple scales consists in some algebraic calculations
that highlight what are the so-called regions of resonance. Indeed the
coefficients of the approximate solution are found to have in the denominators
terms like $n \omega_r-m \omega_{\theta}$.
These $n$ and $m$ cannot take any value, but they depend on the symmetry
of the effective potential, the metric and the perturbation  via  
the  $i,j,k,l$ which are present in the equations : the
allowed $n$ and $m$ indentify  the candidate regions of resonance.
The symmetry restricts further these   regions 
:  it could happen that the constant
 coefficients  cancel and then there would  be no resonance at all.
This calculations performed for a general effective potential
$U_{eff}=U_{eff}(r,\theta)$ showed that the following resonances, usually
present, cannot appear in the {\bf plane-symmetric} case:
$$2:1\:,\:\:3:1\:,\:\:4:1\:,\:\:1:4\:,\:\:1:3\:,\:\:2:3$$
The {\bf only} possible resonances to this degree of approximation are then:
$$1:2\:,\:\:1:1\:,\:\:3:2$$
Since in General Relativity  $\omega_r < \omega_{\theta}$ (this is due
to the curvature of the space), then the only possible resonance is
the $3:2$, which is indeed observed most of the times in black holes candidates.\
 In the next approximation  a new resonance appears,  $4:2$, and 
one may
find many other possible $m:n$. However a property of parametric
resonance is that the region of instability
becomes thinner when $m+n$ increases: the ``higher'' resonances are 
less probable (\cite{bell}). This is why we do not observe them.\\

Landau \& Lifshitz (``Mechanics" 1976), when studying systems similar to
ours (chapter $5$ par.$28$) with the non-geodesic terms $\alpha=0$, use the
method of Lindstedt-Poincar\'e and they  conclude that these are the solution we are interested in, because 
in a closed system, without any source of energy, there cannot be a spontaneus increment in the intensity 
of oscillations. In this sense we asserted that  {\it ``no parametric
resonance occurs for strictly geodesic motion"}(\cite{nostro}): the initial conditions
of the perturbation of the geodesics are related to the energy; even if the
initially small two perturbations couple and
exchange energy, their amplitudes cannot become observable.

\section{$3:2$ resonance near non-rotating compact objects}

An approximate solution for Schwarzschild's perturbed geodesics 
  was obtained and the validity of the approximation was
 verified a posteriori for initial
conditions of the order of $10^{-1}$(which were used in the numerical
integration).\\
In this solution two modes remain locked, with the two frequencies
near to the $3:2$ ratio: the value of these frequencies is were we would see the
twin peaks in the Fourier power spectra of the observed signal.
The correction to the eigenfrequencies can be written in the following
  form (higher order corrections can be neglected):
\begin{eqnarray}
\omega_r^* &\simeq& \omega_r + \omega_r^{(2)}(r_0,\alpha,\tilde{\alpha}_{\theta}(0),\tilde{\alpha}_r(0))\\
\omega_{\theta}^* &\simeq& \omega_{\theta} + \omega_{\theta}^{(2)}(r_0,\alpha,\tilde{\alpha}_{\theta}(0),\tilde{\alpha}_r(0))\nonumber
\end{eqnarray} 

%%%%%%%%%%%%%%%%%%%%%%%%%%%%%%%%%%%%%

\begin{eqnarray}
\omega_{\theta}^{(2)} && [r_0,\alpha,\tilde{\alpha}_{\theta}(0),\tilde{\alpha}_r(0)
] =\nonumber \\ & & \frac{1}{8 \omega_r^2 \omega_{\theta} 
\{\omega_r^2- \omega_{\theta}^2)}\{2 [ c_{11}^2
 \omega_r^2 - \tilde{c}_{21} \omega_r^4 + \nonumber \\
&+&  4 \tilde{c}_{21} \omega_r^2 \omega_{\theta}^2+ 
c_b^2 \omega_r^4 \omega_{\theta}^2 + c_{11}(- e_{20} \omega_r^2 + \nonumber\\&+&
 c_b
  \omega_r^4 + 4 e_{20} \omega_{\theta}^2 - e_{r2} \omega_r^4
+
 4 e_{r2} \omega_r^2
 \omega_{\theta}^2 )] \nonumber \\& & \tilde{\alpha}_r^2(0)
+ [c_{11}(-3 e_{02} \omega_r^2 +8 e_{02} \omega_{\theta}^2 +\nonumber \\
&-& \tilde{e}_{z2} \omega_r^2 \omega_{\theta}^2
+ 8 \tilde{e}_{z2} \omega_{\theta}^4 )+ \omega_r^2 (-3 \tilde{c}_{03} \omega_r^2 + \nonumber \\
&+& 12 \tilde{c}_{03} \omega_{\theta}^2 - 2 c_b e_{02} \omega_{\theta}^2 
+ 2 c_b \tilde{e}_{z2} \omega_{\theta}^4)] \tilde{\alpha}_{\theta}^2(0)\}\:\:, 
\end{eqnarray}

\begin{eqnarray}
\omega_{r}^{(2)} && [r_0,\alpha,\tilde{\alpha}_{\theta}(0),\tilde{\alpha}_r(0)
] = \nonumber \\ & & \frac{1}{24 \omega_r^3  
 (\omega_r^2 - 4 \omega_{\theta}^2)}\{- (\omega_r^2 -4
\omega_{\theta}^2)\nonumber \\&[& 10  e_{20}^2 + 9  e_{30}  \omega_r^2 +
 10  e_{20}  e_{r2}  \omega_r^2 + \nonumber \\ &+&  \omega_r^4 (  3
e_{1re2} + 4 e_{r2}^2 )]  \tilde{\alpha}_r^2(0) + \nonumber \\&+&
6 [-e_{12}\omega_r^4 +4 e_{12} \omega_r^2 \omega_{\theta}^2 +
\nonumber \\ &-& \tilde{e}_{1ze2}\omega_r^4 \omega_{\theta}^2 
+ 4 \tilde{e}_{1ze2}\omega_r^2 \omega_{\theta}^4 + \nonumber \\ &+&  e_{02} (-2
e_{20}\omega_r^2 + 8 e_{20} \omega_{\theta}^2 + \nonumber \\
&+& 4 c_b \omega_r^2 \omega_{\theta}^2) -2 e_{20} \tilde{e}_{z2} \omega_r^2
\omega_{\theta}^2 + \nonumber \\
&-& 2 c_b \omega_{\theta}^2 \omega_r^4 + 8 e _{20} \tilde{e}_{z2} \omega_{\theta}^4
+ 4 c_b \tilde{e}_{z2} \omega_r^2 \omega_{\theta}^4 + \nonumber \\
&+& 2 c_{11} \omega_r^2 (e_{02} - \omega_{\theta}^2 \tilde{e}_{z2})]
\tilde{\alpha}_{\theta}^2(0)\}\:\:.
\end{eqnarray}

%%%%%%%%%%%%%%%%%%%%%%%%%%%%%%%%%%%%%

The dependence on $\alpha$ and $r_0$ is included in the constant coefficients:
 one can vary the initial conditions ($\tilde{\alpha}_{\theta}(0)=
 \epsilon \alpha_{\theta}(0)\sim z(0)$ ,$\tilde{\alpha}_r(0)=
 \epsilon \alpha_r (0)\sim \rho(0)$ ) and/or $\alpha$ \footnote{The
 perturbation $\alpha$
 is included in the coefficients with tilde in the following way:
$$\tilde{c}_{ij}= c_{ij}(1-\alpha)\:\:\:\:.$$
$\alpha_{\theta}(0)$ and $\alpha_r(0)$ ($g$ and $a$ in equations (14))
are the initial values of the amplitudes of the zeroth order
approximation
(the harmonic oscillators).} , obtaining different
values for the corrected frequencies and for their ratio. For small variations
of these parameters, the dependence on them is almost linear.\\
The analytic approximation in figure (\ref{fig3}) was obtained by fixing $|\alpha|=0.99$
(far from geodesics) and $r_0=r_{3:2}$: each point on the segment corresponds to the
state reached by the system starting from differemt initial conditions
($r(0)\in [0.23,0.21]$ and $z(0)\in [-0.26 -0.19 ]$). The same slope
was matched numerically (in the pseudo-Newtonian case) by fixing the initial conditions and by varying $\alpha$\\
The fact that the slope is smaller than $3:2$ depends
on the choice of the initial conditions. This analytic
result is important in explaining the behaviour of the solution more than the numerical
values: the fact that the observed ratios are not exactly in $3:2$
ratio is a feature of parametric resonance and the position of the centroid frequencies
depends on the strength and features of the non-linear perturbation.

%^^^^^^^^^^^^^^^^^^^^^^^^^^^^^^^^^^^^^^^^^^^^^^^^^^^^^^^^^^^^^^^^^^^^^^^

\section{Conclusions}
The Taylor expansion of the relativistic geodesics to the third order leads
to two coupled harmonic oscillators: the purely geodesics motion is
stable, while when it is perturbed there are cases for which parametric
resonance may occur.\\
 Using the method of multiple scales one can highlight
that, owing to the curvature (through the way it determines the 
effective potential), the first allowed resonance between the
radial an the vertical epicyclic frequencies  is the $3:2$, in
agreement with the numerical analysis.\\
Moreover the deviation of the slope from $3:2$ is easily explained
as a property of such a non-linear resonance\\
The analysis of this toy-model reinforces the theory that indeed the
observed pairs of QPOs may be due to parametric resonance, and finally
to the strong gravitational field alone.

\medskip
The work reported here is a part of my 2003 laurea thesis {\it
Risonanza parametrica di dinamiche quasi geodetiche con campi
gravitationali forti} at the Physics Department, University of
Trieste. I thank my external thesis supervisor ({\it correlatore}),
Marek Abramowicz, for suggesting the subject to me, the  many
discussions and his constant support . I also thank my local
supervisor ({\it relatore}), Fabio Benatti, for his advice and help
and Wlodek Klu\'zniak for the help in revising this paper.
 I was working on the problem at the University of Trieste and several
other institutions: at SISSA (Trieste, Italy), Chalmers University
(G\"oteborg, Sweden), UKAFF (Leicester University, England), and the
Cargese 2003 Spring School on Black Holes (Cargese, France). I thank
all these institutions for hospitality.
This work have been supported by the  EU grant in Leicester,by MA's VR
Swedish goverment grant and by the Polish KBN Grant 2P03D01424.

\appendix
\section{Errata corrige}
The equations studied in (\cite{nostro}) for the perturbed geodesics in the
Pseudo-Newtonian potential were incorrect. The correct ones are:

\begin{eqnarray}
f(\rho,z,r_0,\theta_0) &=& c_{11} z \rho + c_b \dot{z} \dot{\rho} + \\
& &c_{21} \rho^2 z+ c_{1b} \rho \dot{z} \dot{\rho} + c_{03} z^3 \nonumber\\
g(\rho,z,r_0,\theta_0) &=& e_{02} z^2 + e_{20}\rho^2 + e_{z2} \dot{z}^2 + \nonumber\\
& & e_{30} \rho^3 + e_{1ze2} \rho \dot{z}^2 + e_{12} \rho z^2 \:\:\:\:. \nonumber
\end{eqnarray}

This does not affect the numerical results: indeed the symmetry (then the possible resonances)
is the same and the only difference is in the values of the initial conditions
which lead to the resonance.
\section{Coefficients} \label{subsec:notation}
In equations (\ref{eqperturbateGR}) I named the constant coefficients in a way practical to remember their
origin : for example $c_{21}$ is a coefficient in the equation $z$ (it has the letter $c$) and it
 it is the coefficient of $\rho$ at the 2nd power and $z$ at the first power
(numbers $(21)$ ). Every coefficient contains the derivatives of the effective potential and of
the metric at equilibrium. The explicit form of each coefficient is:
\begin{eqnarray}
c_{11} &=& E^2 (\frac{1}{2 r_0^2}\frac{\partial}{\partial{r}}
\ddrv{U}{\theta}-\frac{1}{r_0^3} \ddrv{U}{\theta} )\:\:\:\:,\\
c_b&=&-\frac{2}{r_0}\:\:\:\:,\nonumber \\
c_{21} &=& E^2 (\frac{3}{2 r_0^4} \ddrv{U}{\theta}+
\frac{1}{4 r_0^2}\frac{\partial^2}{\partial r^2}\ddrv{U}{\theta}+ \nonumber \\
& &-\frac{1}{r_0^3}\frac{\partial}{\partial{r}}
\ddrv{U}{\theta})\:\:\:\:, \nonumber\\
c_{1b}&=&\frac{2}{r_0^2}\:\:\:\:, \nonumber\\
c_{03}&=&\frac{E^2}{12 r_0^2}\frac{\partial^4
U}{\partial{\theta^4}}\:\:\:\:, \nonumber
\end{eqnarray}
\begin{eqnarray}
e_{02}&=&E^2 [\frac{1}{4}(1-\frac{1}{r_0})\frac{\partial}{\partial{r}}
\ddrv{U}{\theta}]\:\:\:\:, \nonumber\\
e_{20}&=&E^2[\frac{1}{2 r_0^2} \ddrv{U}{r}+\frac{1}{4}(1-\frac{1}{r_0})\frac{\partial^3
U}{\partial{r^3}} ]\:\:\:\:, \nonumber\\
e_{z2}&=&r_0-1\:\:\:\:, \nonumber\\
e_{30}&=& E^2 [\frac{1}{4 r_0^2}\frac{\partial^3
U}{\partial{r^3}}-\frac{1}{2 r_0^3}\ddrv{U}{r}+\nonumber\\
& &+\frac{1}{12}(1-\frac{1}{r_0})\frac{\partial^4
U}{\partial{r^4}} ]\:\:\:\:, \nonumber\\
e_{1ze2}&=&1\:\:\:\:, \nonumber\\
e_{12}&=& \frac{E^2}{4}[\frac{1}{r_0^2}\frac{\partial}{\partial{r}}
\ddrv{U}{\theta}+\frac{\partial^2}{\partial
r^2}\ddrv{U}{\theta}(1-\frac{1}{r_0}) ]\:\:\:\:, \nonumber\\
e_{r2}&=& -\frac{1}{2 r_0}\frac{1}{r_0-1}\:\:\:\:, \nonumber\\
e_{1re2}&=& \frac{1-2 r_0}{2 r_0^2 (r_0-1)^2 }\:\:\:\:, \nonumber
\end{eqnarray}

where the derivatives of the effective potential are evaluated in the
equilibrium  and $E=-u_t$ is the energy.

%^^^^^^^^^^^^^^^^^^^^^^^^^^^^^^^^^^^^^^^^^^^^^^^^^^^^^^^^^^^^^^^^^^^^^^^

\end{document}